\documentclass[12pt]{article}
\newcommand{\ml}{l\kern-0.035cm\char39\kern-0.03cm}

\newtheorem{defi}{Definition 1.}

\newtheorem{ex}{Example}

\newtheorem{prop2}{Proposition 2.}
\newtheorem{coro2}{Corollary 2. 2.}
\newtheorem{defi2}{Definition 2.}

\newtheorem{ex2}{Example 2.}

\newtheorem{prop3}{Proposition 3.}

\newtheorem{defi3}{Definition 3.}

\newtheorem{ex3}{Example 3.}

\begin{document}

\title{ Marginality in non-compatible random events  }
\author{ O\v lga N\'an\'asiov\'a\footnote{Supported by grant
VEGA 1/8833/02, KEGA 3/0038/02,},  Andrei Yu. Khrennikov\footnote{Supported by the EU Human Potential Programme 
under contract N. HPRN-CT-2002-00279 }\\
Dept. of Math. and Descr. Geom.,\\ Faculty of Civil Engineering, \\
Slovak University of Technology, \\
Radlinsk\'eho 11, 813 68 Bratislava, Slovakia\\
International Center for Mathematical Modeling\\
in Physics and Cognitive Sciences\\
 University of V\"axj\"o, S-35195, Sweden\\
Emails: olga@math.sk, Andrei.Khrennikov@msi.vxu.se }

\maketitle
We present a way of introducing joint distibution function and its marginal distribution functions for
non-compatible observables.  Each such marginal distribution function has the property of
commutativity. Models based on this approach can be used to better explain some 
classical phenomena in
stochastic processes.
\section{Introduction}

Let $(\Omega ,\Im  ,P)$ be a probability space and let $\xi_1,\xi_2,\xi_3$ be random 
variables. Then 
$$F_{\xi_1,\xi_2,\xi_3}=P(\{\omega\in\Omega ;\bigcap_{i=1}^3\xi_i^{-1}(-\infty ,r_i)\})$$
is the distribution function and the marginal distribution function  is defined 
by the following way
$$F_{\xi_1,\xi_2}(r_1,r_2)=\lim_{r_3\to\infty}F_{\xi_1,\xi_2,\xi_3}(r_1,r_2,r_3).$$

From the definition of a
distribution function it follows, that all random variables are simultaneously measurable.
It means, that they  can be observable at the same time. 
\

Let $(\Omega_i ,\Im_i,P_i)$ be probability spaces and $i=1,2,...,n$ be the time coordinate.
Let $\xi_i$ be random variable on the probability space $(\Omega_i ,\Im_i,P_i)$. 
How to define the joint distribution function, now?

In this paper,  we will study such random events, which are not simultaneously measurable.
One of  the approache to this problem is studing an algebraic strucute an orthomodular lattice
 (an OML) \cite{D2},\cite{V1}\footnote{We
also mention
so called contextual probabilistic approach\cite{A1}-\cite{A3}}

\begin{defi}
Let $L$ be a nonempty set endowed with a partial ordering $\leq $.
Let there exists the greatest element $1$ and the smallest element
$0$. We consider  operations supremum $(\vee )$, infimum $\wedge $
(the lattice operations ) and an map $\perp :L\to L$ defined as
follows.

\begin{itemize}
\item[(i)]  For any $\{a_n\}_{n\in\mathcal{A}}\in L$, where $\mathcal{%
A\subset N}$ is finite,
\[
\bigvee_{n\in\mathcal{A}} a_n, \bigwedge_{n\in\mathcal{A}} a_n\in L.
\]

\item[(ii)]  For any $a\in L$ $(a^\bot )^\bot =a$.

\item[(iii)]  If $a\in L$, then $a\vee a^\bot =1$.

\item[(iv)]  If $a,b\in L$ such that $a\le b$, then $b^\bot\le a^\bot $.

\item[(v)]  If $a,b\in L$ such that $a\le b$ then $b=a\vee (a^\bot\wedge b)$
(orthomodular law).
\end{itemize}

Then $(L,0,1,\vee ,\wedge ,\perp )$ is said to be \textit{the
orthomodular lattice} (briefly \textit{ the OML}).
\end{defi}

\vskip 1pc Let $L$ be an OML. Then elements $a,b\in L$ will be
called:

\begin{itemize}
\item  \textit{orthogonal} ($a\bot b$) iff $a\le b^\bot$;

\item  \textit{compatible} ($a\leftrightarrow b$) iff there exist mutually
orthogonal elements $a_1,b_1,c\in L$ such that
\[
a=a_1\vee c\mbox {\hskip 1pc  and\hskip 1pc } b=b_1\vee c.
\]
\end{itemize}

\vskip 1pc If $a_i\in L$ for any $i=1,2,...,n$ and $b\in L$ is such, that $%
b\leftrightarrow a_i $ for all $i$, then $b\leftrightarrow\bigvee_{i=1}^n
a_i $ and
\[
b\wedge (\bigvee_{i=1}^n a_i)=\bigvee_{i=1}^n ( a_i\wedge b)
\]
(\cite{D2},\cite{P1},\cite{V1}). \vskip 1pc

Let $a,b\in L$. It is easy to show, that  $a\leftrightarrow b$  if and only if $a=(a\vee b)\wedge (a\vee b^\perp )$ 
 (distributive law). Moreover,   $L$ is a Boolean algebra if and only if $L$ is distributive.
The well known  example of an OML is the lattice of orthogonal projectors in a Hilbert 
 space.

Let $(\Omega ,\Im ,P)$ be a probability space. Then a  statement $A$ is represented as a  measurable 
subset of $\Omega $ ($A\in \Im )$.  For example, if we say $A$ or $B$ it means $A\cup B$
 and non $A$ it means $A^c$ (the set complement in $\Omega $). 

If a basic structure is an OML, then $a$ and $b$ it means infimum ($a\wedge b$), 
$a$ or $b$ it means  supremum ($a\vee b$) and 
non $a$ it means $a^\perp $. 

If $(\Omega ,\Im ,P)$ is a probability space, then for any $A,B\in \Im $ 
$$A=(A\cap B)\cup (A\cap B^c).$$
If $L$ is an OML, then for any $a,b\in L$ 
$$a\geq (a\wedge b)\vee (a\wedge b^\perp).$$

\begin{ex}  Let $L$ be the Hilbert space $R^2$. Then $1:=R^2$ and $0:=[0,0]$. If $a\in L-\{1,0\}$, 
then $a$ is a linear subspace of $R^2$, it means that $a$ is a line,
which contains the point $[0,0]$. We can write, that $a:$ $y=k_ax$. Let $a,b\in L$, $a\neq b$. 
If $a:$ $y=k_ax$, $b:$ $y=k_bx$, 
then $a^\perp :$ $y=-\frac{1}{k_a}$, $a\wedge b= [0,0]$ and $a\vee b=R^2$.   
\end{ex}
On an OML we can define  similar notions as on a measurable space $(\Omega ,\Im )$. 

\begin{defi}
A map $m:L\to [0,1] $ such that

\begin{itemize}
\item[(i)]  $m(0)=0$ and $m(1)=1$.

\item[(ii)]  If $a\bot b$ then $m(a\vee b)=m(a)+m(b)$
\end{itemize}

is called \textit{a state} on $L$. \end{defi}
Let $\mathcal{B(R)}$ be a $\sigma$-algebra of Borel sets. A
homomorphism $x:\mathcal{B(R)}\to L$ is called an observable on
$L$. If $x$ is an observable, then $R(x):=\{x(E);\quad
E\in\mathcal{B(\mathcal{R}) \}}$ is called a range of the
observable $x$. It is clear that $R(x)$ is a Boolean algebra
[Var]. A spectrum of an observable $x$ is defined by the following
way: $\sigma (x)=\cap \{E\in\mathcal{B(R);}$ $x(E)=1\}.$ If $g$ is
a real function, then $g\circ x$ is such observable on $L$ that:

\begin{itemize}
\item[(1.)]  $R(g\circ x)\subset R(x)$;

\item[(2.)]  $\sigma (g\circ x)=\{g(t);\quad t\in \sigma (x)\}$;

\item[(3.)]  for any $E\in\mathcal{B(R)}$

$g\circ x(E) = x(\{t\in \sigma (x); g(t)\in E\})$.
\end{itemize}
We say that $x$ and $y$ are compatible ($x\leftrightarrow y$) if
there exists a Boolean sub-algebra  $\mathcal B\subset L$ such
that $R(x)\cup R(y)\subset \mathcal B$. In other words
$x\leftrightarrow y$ if for any $E,F\in\mathcal{B}(\mathcal{R})$,
$x(E)\leftrightarrow y(F)$.

We call an observable $x$ a finite if $\sigma (x)$ is a finite
set. It means, that $\sigma (x)=\{t_i\}_{i=1}^n$, $n\in N$. Let us
denote $\mathcal{O}$ the set of all finite observables on $L$.

A state is an analogical notion to the probability  measure, 
an observable is analogical  to a random variable.
\vskip 1pc 
\section{ s-map}

Let $L$ be an OML. In the papers \cite{N3},\cite{N4} is defined  $s$-map in the following way:
\begin{defi2}[simult][rpesent]
Let $L$ be an OML. The map $p:L^2\to [0,1]$ 
will be called 
\textit{s-map} if the following conditions hold:

\begin{itemize}
\item[(s1)] $p(1,1)=1$;

\item[(s2)]  if there exists  $a\perp b$, 
then $p(a,b)=0$;

\item[(s3)]  if $a\perp b$, then for any $c\in L$, 
\[
p(a\vee b,c)=p(a,c)+p(b,c) 
\]
\[
p(c,a\vee b)=p(c,a)+p(c,b) 
\]
.
\end{itemize}
\end{defi2}
    
The $s$-map allows  us e.g. to define a conditional probability 
for non compatible random events, a joint  distribution, 
a conditional expectation and covariance for 
non compatible observales. Such random events cannot be 
described the classical probabilty theory\cite{K1}.
This problems are studed in for 
example in \cite{N1}-\cite{N4}. 

In this section we will introduce n-dimensional an s-map (briefly an $s_n$-map)
 and we will  show its basic properties.
\begin{defi2}
Let $L$ be an OML. The map $p:L^n\to [0,1]$ 
will be called 
\textit{an $s_n$-map} if the following conditions hold:

\begin{itemize}
\item[(s1)] $p(1,...,1)=1$;

\item[(s2)]  if there exist  $i$, such that  $a_i\perp a_{i+1}$, 
then $p(a_1,...,a_n)=0$;

\item[(s3)]  if $a_i\perp b_i$, then  
\[
p(a_1,...,a_i\vee b_i,...,a_n)=
p(a_1,...,a_i,...,a_n)+p(a_1,..., b_i,...,a_n), 
\]
for $i=1,...,n$.
\end{itemize}
\end{defi2}
\begin{prop2}
Let $L$ be an OML and let $p$ be an $s_n$-map. Then
\begin{itemize}
\item [(1)]
if $a_i\perp a_j$, then $p(a_1,...,a_n)=0$;
\item [(2)]
for any $a\in L$, a map $\nu :L\to [0,1]$, such that $\nu (a):=p(a,...,a)$ is a state on $L$;
\item [(3)]
for any $(a_1,...,a_n)\in L^n$ $p(a_1,...,a_n)\leq \nu (a_i)$ for each $i=1,...,n$;
\item [(4)]
if $a_i\leftrightarrow a_j$, then 
$$p(a_1,...,a_n)=p(a_1,...,a_{i-1},a_i\wedge a_j,...,a_j\wedge a_i,a_{j+1},...,a_n).$$
\end{itemize}
\end{prop2}
 Proof. 
\begin{itemize}
\item [(1)]
It is enought to prove, that $p(a_1,...,a_n)=0$ if $a_1\perp a_n$. Let $(a_1,...,a_n)\in L^n$ and let $a_1\perp a_n$. 
Then 
$$\begin{array}{clcr} 0\leq p(a_1,...,a_n)&\leq 
p(a_1,...,a_{n-1},a_n)+p(a_1,...,a_{n-1}^\perp ,a_n)\\
&=p(a_1,...,a_{n-2},1,a_n)\\
&=p(a_1,...,a_{n-2},a_n,a_n)+p(a_1,...,a_{n-2},a_n^\perp ,a_n)\\
&=p(a_1,...,a_{n-2},a_n,a_n)\leq ...\leq p(a_1,a_n,...,a_n)\\
&=0.\end{array}$$
From this follows, that $p(a_1,...,a_n)=0$.
\item [(2)]
It is clear, that $\nu (0)=0$, and $\nu (1)=1$. Let $a,b\in L$, such that $a\perp b$. Then 
$$\begin{array}{clcr}
\nu (a\vee b)
&=p(a\vee b,...,a\vee b)\\
&= p(a,a\vee b,...,a\vee b)+p(b,a\vee b,...,a\vee b)\\
&= p(a,a,a\vee b,...,a\vee b)+p(a,b,a\vee b,...,a\vee b)+\\
&\quad  p(b,a,a\vee b,...,a\vee b)+p(b,b,a\vee b,...,a\vee b)\\
&= p(a,a,a\vee b,...,a\vee b)+p(b,b,a\vee b,...,a\vee b)=....\\
&=p(a,...,a)+p(b,...,b)\\
&=\nu (a)+\nu (b).
\end{array}$$
From it follows,that $\nu $ is a state on $L$.
\item [(3)]
Let $(a_1,...,a_n\in L^n$. Then for any $i=1,...,n$ we have 
$$p(a_1,...,a_i,...,a_n)\leq p(a_1,...,a_i,...,a_n)+p(a_1^\perp,...,a_i,...,a_n)$$
and so 
$$p(a_1,a_2,...,a_i,...,a_n)\leq p(1,a_2,...,a_i,...,a_n)=p(a_i,a_2,...,a_i,...,a_n).$$
From it follows, that 
$$p(a_i,a_2,...,a_i,...,a_n)\leq p(a_i,1,...,a_i,...,a_n)=p(a_i,a_i,a_3,...,a_i,...,a_n).$$
Hence
$$p(a_1,...,a_n)\leq p(a_i,...,a_i)=\nu (a_i).$$
\item [(4)]
Let $a,b\in L$, such that $a\leftrightarrow b$. Then $a=(a\wedge b)\vee (a\wedge b^\perp ) $ and
 $b=(b\wedge a)\vee (b\wedge a^\perp ) $. Let $(a_1,...,a_n)\in L^n$ and let $a_1\perp a_2$. 
Then 
$$p(a_1,a_2,...,a_n)=p((a_1\wedge a_2)\vee (a_1\wedge a_2\perp ),a_2,...,a_n).$$
From the property(s3) and for the property (1) we get 
$$p(a_1,a_2,...,a_n)=p(a_1\wedge a_2 ,a_2,...,a_n).$$
And hence
$$p(a_1,a_2,a_3,...,a_n)=p(a_1\wedge a_2 ,a_2\wedge a_1,a_3...,a_n).$$
\end{itemize}
(Q.E.D.)

Let $\bar a=(a_1,...,a_n)\in L^n$. Let us denote $\pi (\bar a)$ a permutation of $(a_1,...,a_n)$.
\begin{prop2}
Let $L$ be an OML. Let $p$ be an $s_n$-map and let $(a_1,...,a_n)\in L^n$. 
\begin{itemize} 
\item [(1)] If there exists $i\in \{1,...,n\}$, such that $a_i=1$, then 
$$p(a_1,...,a_n)=p(a_1,...,a_{i-1}.a_j,a_{i+1},...,a_n)$$
for each $j=1,...,n$.
\item [(2)]
If there exist $i\neq j$ such that $a_i=a_j$, then 
$$p(a_1,...,a_n)=p(\pi (a_1,...,a_n)).$$
\item [(3)]
If there exist $i,j$ such that $a_i\leftrightarrow a_j$, then 
$$p(a_1,...,a_n)=p(\pi (a_1,...,a_n)).$$
\end{itemize}
\end{prop2}
Proof. \begin{itemize}
\item [(1)]
Let $a_i=1$ and let $i\neq j$. Then $a_i=a_j\vee a_j^\perp $ and  from  the Proposition 2.1.(1) 
follows that $p(a_i,...,a_{i-1},a_j^\perp ,a_{j+1},...a_n)=0$. From the property (s3) we get 
$$ p(a_1,...,a_{i-1},1,a_{i+1},...a_n)=p(a_1,...,a_{i-1},a_j,a_{i+1},...a_n)+
$$
$$
p(a_1,...,a_{i-1},a_j^\perp ,a_{i+1},...a_n).$$
And so 
 $$ p(a_1,...,a_{i-1},1,a_{i+1},...a_n)=p(a_1,...,a_{i-1},a_j,a_{i+1},...a_n).$$
\item [(2)]
If $n=2$ and $a_1=a_2$ then it is clear that $p(a_1,a_2)=p(a_2,a_1)$. Let $n\geq 3$ and let $1\neq i$ and $i\neq n$.
Let $a_1=a_n=a$. It is enought to prove, that 
$$p(a,a_2,...,a_i,...,a_{n-1},a)=p(a_i,a_2,...,a_{i-1},a,a_{i+1},...,a_{n-1},a).$$  
From the (1) we have
$$p(a,a_2,...,a_i,...,a_{n-1},a)=p(1,a_2,...,a_{i-1},a_i,a_{i+1},...,a_{n-1},a).$$
From it follows, that
$$p(1,a_2,...,a_i,...,a_{n-1},a)=p(a_i,a_2,...,a_{i-1},a_i,a_{i+1},...,a_{n-1},a)$$
and
$$p(a_i,a_2,...,a_i,...,a_{n-1},a)=p(a_i,a_2,...,a_{i-1},1,a_{i+1},...,a_{n-1},a).$$
From the aditivity it follows, that
$$p(a_i,a_2,...,a_{i-1},1,a_{i+1},...,a_{n-1},a)=p(a_i,a_2,...,a_{i-1},a,a_{i+1},...,a_{n-1},a).$$
Hence
$$p(a,a_2,...,a_{i-1},a_i,a_{i+1},...,a_{n-1},a)=p(a_i,a_2,...,a_{i-1},a_1,a_{i+1},...,a_{n-1},a_n).$$
From it follows that $p(a_1,...,a_n)p(\pi (a_1,...,a_n)$, if there exist $i,j$, such that $i\neq j$ and $a_i=a_j$.
\item [(3)]
Let  $a_1\leftrightarrow a_n$. Then 
$$p(a_1,...,a_n)=p(a_1\wedge a_n,a_2,...,a_{n-1},a_n\wedge a_1).$$
Because $a_1\wedge a_n=a_n\wedge a_1$ and from the property (2) it follows, that 
$$p(a_1,...,a_n)=p(\pi (a_1,...,a_n)).$$
\end{itemize}
(Q.E.D.)

Let $\Pi (\bar a)$ be the set of all permutions and let 
$$\bar a^{(i)}_{(k)}=(a_1,...,a_{k-1},a_k,a_{k+1},...,a_{i-1},a_k,a_{i+1},...,a_n).$$ 

\begin{coro2}
Let $L$ be an OML. Let $p$ be an $s_n$-map and let $\bar a\in L^n$. 
\begin{itemize} 
\item [(1)] If there exists $i\in \{1,...,n\}$, such that $a_i=1$, then 
$$p(\bar a)=p(\bar b)$$
for each $\bar b\in\bigcup_{k}\Pi(\bar a^{(i)}_{(k)})$.
\item [(2)]
If there exist $i\neq j$ such that $a_i=a_j$, then 
$$p(\bar a)=p(\bar b) $$
for each $\bar b\in\bigcup_{k}\Pi(\bar a^{(i)}_{(k)})$.
\item [(3)]
If there exist $i,j$ such that $a_i\leftrightarrow a_j$, then 
$$p(\bar a)=p(\bar b)$$
for each $\bar b\in\bigcup_{k}\Pi(\bar a^{(i)}_{(k)})$.
\end{itemize}
\end{coro2}
\begin{ex2}
Let $n=3$ and $a,b\in L$. If $\bar a=(a,a,b)$, then
$$
\Pi (\bar a)=\{(a,a,b),(b,a,a),(a,b,a)\}
$$ 
and 
$\bar a^{(1)}_{(3)}=(b,a,b)$, $\bar a^{(2)}_{(3)}=(a,b,b)$,$\bar a^{(1)}_{(2)}=(a,a,b)$.
Hence 
$$p(a,a,b)=p(a,b,a)=p(b,a,a)=p(b,b,a)=p(a,b,b)=p(b,a,b).$$

Let $n=4$ and $a,b,c\in L$. If $\bar a=(a,b,c,c)$, then $\bar a^{(4)}_{(2)}=(a,b,c,b)$ and
$$p(a,a,b,c)=p(a,b,c,a)=p(b,b,c,a)=...=p(c,a,b,c).$$
\end{ex2} 

\section{The joint distribution function and marginal distribution funtions}

\begin{defi3}
Let $L$ be an OML and let $p$ be an $s_n$-map. If $x_1,...,x_2$ are observables on $L$, then the map 
$$p_{x_1,...,x_n}:\mathcal B(R)^n\to [0,1],$$
such that 
$$p_{x_1,...,x_n}(E_1,...,E_n)=p(x_1(E_1),...,x_n(E_n))$$
is called the joint distribution of the observables $x_1,...,x_n$.
\end{defi3} 

\begin{defi3}
Let $L$ be an OML and let $p$ be an $s_n$-map. If $x_1,...,x_2$ be observables on $L$,  then the map 
$$F_{x_1,...,x_n}:R^n\to [0,1],$$
such that 
$$F_{x_1,...,x_n}(r_1,...,r_n)=p(x_1(-\infty ,r_1),...,x_n(-\infty ,r_n))$$
is called the joint distribution function of the observables $x_1,...,x_n$.
\end{defi3} 
\begin{defi3}
Let $L$ be an OML and let $p$ be an $s_n$-map. If $x_1,...,x_2$ be observables on $L$,  then 
a marginal distribution function is
$$\lim_{x_i\to\infty}F_{x_1,...,x_i,...,x_n}(r_1,...,r_i,...,r_n).$$
\end{defi3} 
\begin{defi3}
Let $L$ be an OML and let $p$ be an $s_n$-map. Let $x_1,...,x_2$ be observables on $L$ and  
$F_{x_1,...,x_n}$ be the joint distribution function of the observables $x_1,...,x_n$.
Then we say, that  $F_{x_1,...,x_n}$ has the property of commutativity if for each $(r_1,...,r_n)\in R^n$ 
$$F_{x_1,...,x_n}(r_1,...,r_n)=F_{\pi (x_1,...,x_n)}(\pi (r_1,...,r_n)).$$ 
\end{defi3} 

It is clear that $F_{x_1,...,x_n}$ has the property of commutativity if and only if 
$$p(x_1(E_1),...,x_n(E_n))=p(\pi (x_1(E_1),...,x_n(E_n))),$$
for each $E_i\in\mathcal B(R)$, $i=1,...,n$.

\begin{prop3}
Let $L$ be an OML and let $p$ be an $s_n$-map. Let $x_1,...,x_2\in\mathcal O$ and let   
$F_{x_1,...,x_n}(r_1,...,r_n)$
be the joint distribution function of the observables $x_1,...,x_n$.
\begin{itemize}
\item [(1)]
For each $(r_1,...,r_n)\in R^n$ $0\leq F_{x_1,...,x_n}(r_1,...,r_n)\leq 1$;
\item [(2)]
If $r_i\leq s_i$, then  $F_{x_1,...,x_n}(r_1,...,r_i,...,r_n)\leq F_{x_1,...,x_n}(r_1,...,s_i,....,r_n)$.
\item[(3)]
For each $i=1,...,n$
$$\lim_{r_i\to\infty }F_{x_1,...,x_n}(r_1,...,r_n)=F_{x_1,...,x_n}(r_1,...r_{i-1},1,r_{i+1},...,r_n).$$
\item[(4)]
For each $i=1,...,n$
$$\lim_{r_i\to -\infty }F_{x_1,...,x_n}(r_1,...,r_n)=0.$$
\item[(5)]
If there exist $i,j$, such that $i\neq j$ and $x_i \leftrightarrow x_j$, then 
$$F_{x_1,...,x_n}(r_1,...,r_n)=F_{\pi (x_1,...,x_n)}(\pi (r_1,...,r_n)).$$
\end{itemize}
\end{prop3}
Proof.
\begin{itemize}
\item [(1)] It follows directly from the definition of the function $F_{x_1,...,x_n}$.
\item [(2)] 
Let $r_i\leq s_i$. Then $(-\infty ,r_i)\subseteq  (-\infty ,s_i)$ and so $x_i((-\infty ,r_i))\leq x_i((-\infty ,r_i))$
and $x_i((-\infty ,s_i))= x_i((-\infty ,r_i))\vee x_i([r_i,s_i))$. From it follows, that 
$$
F_{x_1,...,x_n}(r_1,...,s_i,...,r_n)=
$$
$$
p(x_1((-\infty ,r_1)),...,x_i((-\infty ,r_i)),....,x_n((-\infty ,r_n))+
$$
$$
\quad p(x_1((-\infty ,r_1)),....,x_i([r_i,s_)),....,x_n((-\infty ,r_n))
$$
$$
=F_{x_1,...,x_n}(r_1,...,r_i,...,r_n)+
$$
$$
\quad p(x_1((-\infty ,r_1)),....,x_i([r_i,s_)),....,x_n((-\infty ,r_n))
$$
and so 
$$ F_{x_1,...,x_n}(r_1,...,s_i,...,r_n)\geq F_{x_1,...,x_n}(r_1,...,r_i,...,r_n).$$
\item [(3)]
Because $x_i\in\mathcal O$, then there exist $r_{i0}\in R$, such that
for any $r\geq r_{i0}$ $\sigma (x_i)\subseteq (-\infty ,r)$ and so 
$x_i(-\infty ,r))=1$.
Hence 
$$\lim_{r_i\to\infty }F_{x_1,...,x_n}(r_1,...,r_n)=F_{x_1,...,x_n}(r_1,...r_{i-1},1,r_{i+1},...,r_n).$$
\item [(4)]
Because $x_i\in\mathcal O$, then there exist $r_{i0}\in R$, such that for each $r\leq r_{i0}$  
  $(-\infty ,r)\cap\sigma (x_i)=\emptyset$  and so 
$x_i(-\infty ,r))=0$.
Hence 
$$\lim_{r_i\to -\infty }F_{x_1,...,x_n}(r_1,...,r_n)=0.$$
\item [(5)]
Because $F_{x_1,...,x_n}(r_1,...,r_n)=p(x_1((-\infty, r_1)),...,x_n(((-\infty ,r_n))$, then it 
follows directly from the Proposition 2.2.
\end{itemize}
(Q.E.D.)
\begin{prop3}
Let $L$ be an OML and let $p$ be an $s_n$-map. Let $x_1,...,x_n\in\mathcal O$ and let $F_{x_1,...,x_n}(r_1,...,r_n)$ be 
the joint distribution function. Compatibility of just two observables imply the total 
commutativity.
\end{prop3}
Proof. It follows directly from the definition of the joint distribution function and from the Proposition 2.2.

\begin{prop3}
Let $L$ be an OML and let $x_1,...,x_n\in \mathcal O$. Then there exist a probability space $(\Omega ,\Im , P)$ and 
random variables $\xi_1,...,\xi_n$ on it, such that  
$$F_{x_1,...,x_n}(r_1,...,r_n)=F_{\xi_1,...,\xi_n}(r_1,...,r_n)$$
and $P_{\xi_i}$ such that 
$$P_{\xi_i}((-\infty ,r))=\nu (x_i(-\infty ,r)),$$
where $r\in R$ and $i=1,...,n$ is the probability distribution of the random varaible $\xi_i$, 
\end{prop3}
Proof.
Let $\Omega =\sigma (x_1)\times ...\times \sigma (x_n)$ and 
let $\Im =2^\Omega $. Then  each $\omega =(r_1,...,r_2)$ $\xi_i(\omega_1,...,\omega_n)=\omega_i$. 
Let $A\subset \Im $ and let $P:\Im \to [0,1]$, such that 
$$P(A)=\sum_{\omega\in A}p(x_1(\xi_1(\omega )),...,x_n(\xi_n(\omega ))).$$
$$F_{x_1,...,x_n}(r_1,...,r_n)=F_{\xi_1,...,\xi_n}(r_1,...,r_n).$$
It is clear, that $P(\emptyset )=0$ and $P(\Omega )=P(\sigma (x_1)\times \sigma (x_n))=1$.  
Let $A,B\in\Im $, such that $A\cap B=\emptyset$. Then 
$$ P(A\cup B)=\sum_{\omega\in A\cup B}p(x_1(\xi_1(\omega )),...,x_n(\xi_n(\omega )))$$
and so 
$$P(A\cup B)=\sum_{\omega\in A}p(x_1(\xi_1(\omega )),...,x_n(\xi_n(\omega )))
+\sum_{\omega\in B}p(x_1(\xi_1(\omega )),...,x_n(\xi_n(\omega ))).$$
From it follows that 
$$P(A\cup B)=P(A)+P(B).$$
From the fact, taht $\Omega $ is the finite set follows, that $P$ is the $\sigma $-aditive measure  and so 
$(\Omega ,\Im, P)$ has the same properties as a classical probability space and $\xi_i:\Omega\to R$ is a measurable 
function on it. For each  $r\in R$ 
 $$P_{\xi_1}((-\infty ,r))=P(\xi_1^{-1}(-\infty ,r))=P((-\infty ,r)\times\sigma (x_2)\times ... \times \sigma (x_n))$$
and then
$$ P_{\xi_1}(-\infty ,r)=\sum_{\omega\in (-\infty ,r) \times\sigma (x_2)\times ... \times \sigma (x_n)}
p(x_1(\xi_1(\omega )),...,x_n(\xi_n(\omega))).$$
From it follows, that 
$$P_{\xi_1}(-\infty ,r)=p(x_1((-\infty ,r)),1,...,1)=\nu (x_1(-\infty ,r)).$$
From the defnition of the marginal distributiuon function it follows, that 
$$\nu (x_i(-\infty ,r))=F_{\xi_i }(r)$$
is the distribution function for the observable $\xi_i$ and
$$p_{x_1,...,x_n}(r_1,...,r_n)=F_{\xi_1,...,\xi_n}(r_1,...,r_n).$$
is a joint distribution for that vector of random variables $(\xi_1,...,\xi_n)$.
(Q.E.D.)

If we consider a quantum model as  an OML, a marginal distribution function
 defined by using an $s_n$-map has the property of commutativity.  
It follows that, in general, it need not true that 
that   
$$F_{x_1,...,x_n}(t_1,...,t_n)=F_{x_1,...,x_{n+1}}(t_1,...,t_n,\infty ),$$  
where $F_{x_1,...,x_n}(t_1,...,t_n)$, $F_{x_1,...,x_{n+1}}(t_1,...,t_n,\infty )$ 
are  joint distribution funtions and $x_1,...,x_{n+1}$ are  observables 
on $L$. Consequently,  we can find such an $s_n$-map  and an $s_{n+1}$-map such 
that $$p(a_1,...,a_n)\neq p(a_1,...,a_n,1)$$ on $L$.  
Moreover if $$p(a_1,...,a_n)= p(a_1,...,a_n,1)$$ on $L$, then the
 $s_n$-map has the property of commutativity. This is not true in general, 
 either 
(\cite{N3},\cite{N4}).
\begin{ex3}
Let $L=\{a,a^\perp ,b,b^\perp ,c,c^\perp ,0,1\}$. Let $x,y,z\in\mathcal O$. Let 
$\sigma (x)=\sigma (y)=\sigma (z)=\{-1,1\}$. Let $x(1)=a$, $x(1)=b$ and $z(1)=c$.
Let an $s_3$-map be defined by the following way:
$$p(a,a,a)=0.3,\quad p(b,b,b)=0.4,\quad p(c,c,c)=0.5,$$
$$p(a,b,1)=0.1,\quad p(a,b^\perp ,1)=0.2,\quad p(a^\perp ,b,1)=0.3,\quad p(a^\perp ,b^\perp ,1)=0.4, $$
$$p(a,c,1)=0.2,\quad p(a,c^\perp ,1)=0.1,\quad p(a^\perp ,c,1)=0.3,\quad p(a^\perp ,c^\perp ,1)=0.4, $$
$$p(b,c,1)=0.2,\quad p(b,c^\perp ,1)=0.2,\quad p(b^\perp ,c,1)=0.3,\quad p(b^\perp ,c^\perp )=0.3, $$
$$p(a,b,c)=0,\quad p(a,b, c^\perp )=0.1,\quad p(a,b^\perp ,c)=0.2,\quad p(a,b^\perp ,c^\perp )=0, $$
$$p(a^\perp ,b,c)=0.2,\quad p(a^\perp ,b, c^\perp )=0.1,
$$
$$
\quad p(a^\perp ,b^\perp ,c)=0.1,\quad p(a^\perp ,b^\perp ,c^\perp ,1)=0.3,$$
$$p(b,a,c)=0.1,\quad p(b,a, c^\perp )=0,\quad p(b^\perp ,a,c)=0.1,\quad p(b^\perp ,a,c^\perp )=0.1, $$
$$p(b,a^\perp ,c)=0.1,\quad p(b,a^\perp , c^\perp )=0.2,
$$
$$
\quad p(b^\perp ,a^\perp ,c)=0.2,\quad p(b^\perp ,a^\perp ,c^\perp ,1)=0.2,$$
$$p(c,a,b)=0.01,\quad p(c,a, b^\perp )=0.19,\quad p(c,a^\perp ,b )=0.19,\quad p(c ,a^\perp ,c^\perp )=0.11, $$
$$p(c^\perp ,a,b)=0.09,\quad p(c^\perp ,a, b^\perp )=0.01,
$$
$$
\quad p(c^\perp ,a^\perp ,b)=0.11,\quad p(c^\perp ,a^\perp ,b^\perp ,1)=0.29,$$
$$
p(a,b,c)=p(a,c,b),\quad p(b,a,c)=p(b,c,a),\quad p(c,a,b)=p(c,b,a)$$
$$.................................................................$$
$$p(a^\perp ,b^\perp ,c^\perp)=p(a^\perp,c^\perp,b^\perp),\quad p(b^\perp,a^\perp,c^\perp)=
$$
$$
p(b^\perp, c^\perp, a^\perp),\quad p(c^\perp, a^\perp, b^\perp)=p(c^\perp, b^\perp, a^\perp).$$
Then $p$ is  an $s_3$-map and 
$$F_{x_1,x_2,x_3}(1,1,1)=p(a^\perp ,b^\perp ,c^\perp)=0.3,$$
$$F_{x_2,x_1,x_3}(1,1,1)=p(b^\perp ,a^\perp ,c^\perp)=0.2,$$
$$F_{x_3,x_2,x_1}(1,1,1)=p(c^\perp ,b^\perp ,a^\perp)=0.29,$$
$$
\lim_{r_1\to\infty }  F_{x_1,x_2,x_3}(r_1,r_2,r_3)=
$$
$$
p(1 ,y(r_2) ,z(r_3))=p(1 ,z(r_3) ,z(r_2))
=
\lim_{r_1\to\infty }  F_{x_1,x_3,x_2}(r_1,r_3,r_2),$$
where $r_2,r_3\in R$.
\end{ex3}

\end{document}